\documentclass[acus]{epac98}

\usepackage{epsfig}

\newcommand{\bit}{\begin{Itemize}}
\newcommand{\eit}{\end{Itemize}}

\begin{document}
\title{DISCUSSION ON MUON COLLIDER PARAMETERS AT CENTER OF MASS ENERGIES
FROM 0.1 TEV TO 100 TEV}
\author{ Bruce J. King, Brookhaven National Laboratory $^1$ }
\maketitle

\begin{abstract}
 Some of the potential capabilities and design challenges of muon colliders
are illustrated using self-consistent collider parameter sets at center
of mass energies ranging from 0.1 TeV to 100 TeV.

\end{abstract}

\footnotetext[1]{
web page: http://pubweb.bnl.gov/people/bking/,
email: bking@bnl.gov.
This work was performed under the auspices of
the U.S. Department of Energy under contract no. DE-AC02-98CH10886. }



\section{Introduction}
\label{sec-intro}

 The main motivation for research and development efforts on muon collider
technology is the assertion that affordably priced muon colliders might
provide lepton-lepton collisions at much higher center of mass (CoM) energies
than is feasible
for electron colliders, and perhaps eventually explore the spectrum of
elementary particles at mass scales inaccessible even to hadron colliders.

 This paper attempts to present some justification for these assertions
through discussion and evaluation of the self-consistent muon
collider parameter sets given in table 1, at CoM energies ranging from 0.1
to 100 TeV.

  The parameter set at 0.1 TeV CoM energy was included as a lower
energy reference point and was constrained to essentially reproduce one of
the sets of parameters currently under study[1] by the Muon Collider
Collaboration (MCC). In contrast, the other parameter sets represent
speculation by the author on how the parameters might evolve with CoM energy
and they have not been studied or discussed in detail within the MCC.

\section{Generation of Parameter Sets}
\label{sec-how}

 The parameter sets in table 1 were generated through iterative runs of a
stand-alone FORTRAN program, LUMCALC. The parameter sets are calculated
from the input values for several input parameters --
namely, the CoM energy (${\rm E_{CoM}}$), the collider ring circumference (C)
and depth below the
Earth's surface (D), the beam momentum spread ($\delta$) and 6-dimensional
invariant
emittance (${\rm \epsilon_{6N}}$), the reference pole-tip magnetic field
for the final focus quadrupoles (${\rm B_{4\sigma}}$), and the
time until the beams are dumped (${\rm t_D}$) -- and from the input of
maximum allowable
values for several other parameters -- namely, the bunch repetition
frequency (${\rm f_b}$),
the initial number of muons per bunch ($N_0$), the beam-beam tune disruption
parameter ($\Delta \nu$),
the beam divergence at the interaction point ($\sigma_\theta$), the
maximum aperture for the final focus quadrupoles (${\rm A_{\pm4\sigma}}$), and
maximum allowable neutrino radiation where the plane of the collider ring
cuts the Earth's surface.

 As a preliminary stage of calculation, LUMCALC makes any parameter
adjustments that may be required to satisfy the input constraints. These
are, in order:
1) reducing $\sigma_\theta$ to the limit
 imposed by ${\rm A_{\pm4\sigma}}$ (based on scaling to existing final focus
designs at 0.1 TeV and 4 TeV[1]),
2) reducing ${\rm N_0}$ to
 attain an acceptable $\Delta \nu$, and
3) reducing ${\rm f_b}$ until
 the neutrino radiation is acceptable.

The output luminosity may be derived in terms of the input parameters as:
\begin{eqnarray}
\cal{L}
{\rm   [cm^{-2}.s^{-1}] }
   & = & {\rm 2.11 \times 10^{33} \times H_B
                   \times (1-e^{-2t_D[\gamma \tau_\mu]}) }
                                                           \nonumber \\
   & &  \times {\rm  \frac{ f_b[s^{-1}] (N_0[10^{12}])^2 (E_{CoM}[TeV])^3}
                          { C[km] } } \nonumber \\
   & & {\rm \times \left( \frac{\sigma_\theta [mr].\delta[10^{-3}]}
                   {\epsilon_{6N}[10^{-12}]}  \right) ^{2/3}  }.
\end{eqnarray}
This formula uses the standard MCC assumption[1] that
the ratio of transverse to longitudinal emittances can be chosen
in the muon cooling channel to maximize the luminosity for a
given ${\rm \epsilon_{6N}}$. The pinch enhancement factor, ${\rm H_B}$,
is very close to unity (see table 1), and the numerical coefficient
in equation 1 includes a geometric correction factor of 0.76 for the
non-zero bunch length,
$\sigma_z = \beta^*$ (the ``hourglass effect'') .

\section{Discussion}
\label{sec-discuss}

  The physics motivation for each of the parameter sets in table 1 is
discussed in [2]. Briefly, the number of $\mu\mu \rightarrow {\rm ee}$
events gives
a benchmark estimate of the discovery potential for elementary particles
at the full CoM energy of the collider, while the production of hypothesized
100 GeV Higgs particles indicates roughly how the colliders would perform in
studying physics at this fixed energy scale.

 Further information on the important issue of neutrino radiation
can be found in [3]. The numbers given in table 1 come from an
analytical calculation that is not intended to be accurate at much better
than an order of magnitude level and that is deliberately conservative,
i.e. it may well overestimate the radiation levels. The radiation levels
are predicted to rise approximately as the cube of the collider energy
if other relevant parameters are held fixed (up to some mitigating factors
that come into play at the highest energies), rapidly becoming a serious
design constraint for colliders at the TeV scale and above.

  The 1 TeV parameter set of table 1 would give about the same luminosity as,
for example, the design for the proposed NLC linear electron collider at
the same energy, and the physics motivation and capabilities might be
relatively similar[2,4]. Placement of the collider at 125 meters depth
reduces the average neutrino radiation in the collider plane to
less than one thousandth of the U.S.
federal off-site radiation limit (1 mSv/year, which is of the same order of
magnitude as the typical background radiation from natural causes).
Nevertheless, attention would still need to be paid to minimizing
the length, L, of any straight sections with low beam divergence, since
these produce radiation hotspots with intensity proportional to L[3].

 The 4 TeV parameter set was chosen as being at about the highest energy
that is practical for a ``first generation'' muon collider on an existing
laboratory site, due to neutrino radiation, and the muon current has been
reduced to lower the radiation to the same level as the 1 TeV parameter
set, accepting the consequent loss in luminosity.

 The 4 TeV parameters may be compared to the MCC 4 TeV design presented
at Snowmass'96[5], which did not take account of the neutrino
radiation issue and hence attained a luminosity higher by more than an order
of magnitude. The lower bunch repetition rate of the current 4 TeV parameter
set makes some of the design parameters more relaxed than in the Snowmass
design, particularly in allowing a
``lite front end'' with relaxed rate specifications: the
proton driver, pion production target and cooling channel.
On the other hand, the desire to recover some of the lost luminosity motivates
collider ring parameters that are slightly more aggressive,
especially $\beta^*$ (3 mm reduced to 1.2 mm) and $\sigma_\theta$
(0.9 mrad increased to 1.6 mrad). This entails a more difficult
final focus design and also a more difficult task to shield the detector
region from muon decay backgrounds.

 Beyond CoM energies of a few TeV, it is probably necessary to build
the colliders at isolated sites where the public would not be exposed to
the neutrino radiation disk.
These will presumably be
``second generation'' machines, arriving after the technology of muon
colliders has been established in one or more smaller and less expensive
machines built at existing HEP laboratories.
The gain from being able to relax the neutrino radiation constraint is 
evident in the 10 TeV parameter set, with an exciting luminosity of
$1.0 \times 10^{36} {\rm cm^{-2}.s^{-1}}$ at several times the
discovery mass reach of the LHC hadron collider.

 From the progression of the parameter sets it is clear that the final
focus design will become progressively more difficult with rising CoM
energy. Consider, for example, the overall beam demagnification,
$\sqrt{\beta_{\rm max}/\beta^*}$,
a dimensionless parameter that should be closely correlated with
fractional tolerances in magnet uniformity, residual chromaticity
etc. For the 10 TeV example, this has risen to approximately 31 000
in both the x and y coordinates,
which -- as a very crude comparison that ignores considerable differences
in the other final focus parameters -- happens to be approaching the
y-coordinate value for both the 0.5 TeV (IA) and 1.0 TeV (IIA) designs
for the NLC linear collider (i.e. 39 000, with
${\rm \beta_{max, y} = 190\; km}$ and ${\rm {\beta^*}_y = 0.125\; mm}$) [4].
The spot size -- clearly indicative of
vibration and alignment tolerances -- is also falling, but even at 100 TeV
it remains an order of magnitude above the spot size in the
y coordinate for the NLC design parameters.
For perspective, then, the design of the final focus at 10 TeV CoM energy
may well still be less challenging than the design of the muon cooling channel,
and the latter task is essentially independent of the collider energy
(up to assumed advances for later generation colliders).

 The highest energy parameter set in table 1, at 100 TeV, clearly presents
the most difficult design challenge, for several reasons: 1) cost reductions
will be needed to make a machine of this size affordable,
2) siting will
be more difficult than at 10 TeV, since the neutrino radiation is now
well above the U.S. federal limit, 
3) $\sqrt{\beta_{\rm max}/\beta^*}$ is almost an order of magnitude
larger than at 10 TeV,
4) The assumed $\epsilon_{6N}$ is 25 times smaller than for
the 10 TeV parameters, albeit with much smaller bunches, so the assumed phase
space density is nearly a factor of two larger,
and finally
5) the beam power has risen to 170 MW, with synchrotron radiation
rising rapidly to contribute a further 110 MW.

 Most of these extrapolations correspond to incremental advances in
technology, particularly involving magnets: magnetic field strength (for
improved cooling and final focus, smaller accelerating rings and collider
rings), stability and uniformity (particularly for the final focus) and cost
reduction (for the accelerating rings and collider rings). Hence, it is
certainly not ruled out that such a parameter set could become achievable
after a couple of decades of research and development dedicated to muon
collider technology.

\section{Conclusions}
\label{sec-conc}

 It has been shown that muon collider parameter sets at up to 10 TeV CoM
energy may well be realistic by today's standards of technology while muon
colliders at the 100 TeV energy scale require technological extrapolations
that could perhaps be achievable within the relatively near-term future.


\section{references}


[1] The Muon Collider Collaboration, ``Status of Muon Collider Research
and Development and Future Plans'', to be submitted to Phys. Rev. E.

\noindent [2] B.J. King, ``Muon Colliders: New Prospects for Precision
Physics and the High Energy Frontier'',
BNL CAP-224-MUON-98C,
submitted to Proc. Latin Am. Symp.
on HEP, April 8-11,1998, San Juan, Puerto Rico, Ed. J.F. Nieves.
Available at http://pubweb.bnl.gov/people/bking/. 

\noindent [3]  B.J. King,
    ``A Characterization of the Neutrino-Induced
    Radiation Hazard at TeV-Scale Muon Colliders'',
    BNL CAP-162-MUON-97R,
    to be submitted for publication.

\noindent [4] The NLC Design Group, ``Zeroth-Order Design Report for the
   Next Linear Collider'', LBNL-PUB-5424, SLAC Report 474,
   UCRL-ID-124161, May 1996.

\noindent [5] The Muon Collider Collaboration,
``$\mu^+\mu^-$ Collider: A Feasibility Study'', BNL-52503,
Fermilab-Conf-96/092, LBNL-38946, July 1996.
\newpage

\begin{table}[htb!]
\begin{center}
\caption{Self-consistent parameter sets for muon colliders. The generation of
these parameter sets is discussed in the text. Except for the first parameter
set, which has been studied in some detail by the Muon Collider Collaboration,
these parameters represent speculation by the author on how muon colliders
might evolve with energy. The beam parameters at the interaction point are
defined to be equal in the horizontal (x) and vertical (y) transverse
coordinates.}
\begin{tabular}{|r|ccccc|}
\hline
\multicolumn{1}{|c|}{ {\bf center of mass energy, ${\rm E_{CoM}}$} }
                            & 0.1 TeV & 1 TeV & 4 TeV &  10 TeV  & 100 TeV \\
\multicolumn{1}{|c|}{ {\bf description} }
                            & MCC para. set & LHC complement & E frontier &
                                        $2^{\rm nd}$ gen. & ult. E scale \\
\hline \hline
\multicolumn{1}{|l|}{\bf collider physics parameters:} & & & &  & \\
luminosity, ${\cal L}$ [${\rm cm^{-2}.s^{-1}}$]
                                        & $1.2 \times 10^{32}$
                                        & $1.0 \times 10^{34}$
                                        & $6.2 \times 10^{33}$
                                        & $1.0 \times 10^{36}$
                                        & $4.0 \times 10^{36}$ \\
$\int {\cal L}$dt [${\rm fb^{-1}/det/year}$]
                                        & 1.2 & 100 & 62 & 10 000
                                        & 40 000 \\
No. of $\mu\mu \rightarrow {\rm ee}$ events/det/year
                                        & 10 000 & 8700 & 340 & 8700 & 350 \\
No. of 100 GeV SM Higgs/det/year        & 1600 & 69 000
                                         & 69 000 & $1.4 \times 10^7$
                                         & $8.3 \times 10^7$ \\
fract. CoM energy spread, ${\rm \sigma_E/E}$ [$10^{-3}$]
                                        & 0.85 & 1.6 & 1.6 & 1.0 & 1.0 \\
\hline
\multicolumn{1}{|l|}{\bf collider ring parameters:}   & & & & & \\
circumference, C [km]                   & 0.3 & 2.0 & 7.0 & 15 & 100 \\
ave. bending B field [T]               & 3.5 & 5.2 & 6.0 & 7.0 & 10.5 \\
\hline
\multicolumn{1}{|l|}{\bf beam parameters:}            & & & & & \\
($\mu^-$ or) $\mu^+$/bunch,${\rm N_0[10^{12}}]$
                                        & 4.0 & 3.5 & 3.1 & 2.4 & 0.18 \\
($\mu^-$ or) $\mu^+$ bunch rep. rate, ${\rm f_b}$ [Hz]
                                        & 15 & 15 & 0.67 & 15 & 60 \\
6-dim. norm. emittance, $\epsilon_{6N}
               [10^{-12}{\rm m}^3$]    & 170 & 170 & 170 & 50 & 2 \\
x,y emit. (unnorm.)
              [${\rm \pi.\mu m.mrad}$] & 210 & 12 & 3.0 & 0.55 & 0.0041 \\
x,y normalized emit.
              [${\rm \pi.mm.mrad}$]    & 99 & 57 & 57 & 26 & 1.9 \\
fract. mom. spread, $\delta$ [$10^{-3}$]
                                       & 1.2 & 2.3 & 2.3 & 1.4 & 1.4 \\
relativistic $\gamma$ factor, ${\rm E_\mu/m_\mu}$
                                        & 473 & 4732 & 18 929 & 47 322
                                        & 473 220 \\
ave. current [mA]                      & 20 & 10 & 0.46 & 24 & 4.2 \\
beam power [MW]                        & 1.0 & 8.4 & 1.3 & 58 & 170 \\
decay power into magnet liner [kW/m]   & 1.1 & 0.58 & 0.03 & 1.4 & 1.3 \\
time to beam dump,
          ${\rm t_D} [\gamma \tau_\mu]$ & no dump & 0.5 & 0.5 & no dump & 0.5 \\
effective turns/bunch                  & 519 & 493 & 563 & 1039 & 985 \\
\hline
\multicolumn{1}{|l|}{\bf interaction point parameters:}      & & & & & \\
spot size, $\sigma_x = \sigma_y
                          [\mu {\rm m}]$   & 80 & 7.6 & 1.9 & 0.78 & 0.057 \\
bunch length, $\sigma_z$ [mm]          & 31 & 4.7 & 1.2 & 1.1 & 0.79 \\
$\beta^*$ [mm]                          & 31 & 4.7 & 1.2 & 1.1 & 0.79 \\
ang. divergence, $\sigma_\theta$
                             [mrad]    & 2.6 & 1.6 & 1.6 & 0.71 & 0.072 \\
beam-beam tune disruption parameter, $\Delta \nu$
                                        & 0.044 & 0.066 & 0.059 & 0.100
                                        & 0.100 \\
pinch enhancement factor, ${\rm H_B}$  & 1.007 & 1.040 & 1.025 & 1.108
                                        & 1.134 \\
beamstrahlung fract. E loss/collision  & $2.1 \times 10^{-14}$
                                       & $1.2 \times 10^{-10}$
                                       & $2.3 \times 10^{-8}$
                                       & $2.3 \times 10^{-7}$
                                       & $3.2 \times 10^{-6}$ \\
\hline
\multicolumn{1}{|l|}{\bf final focus lattice parameters:} & & & & & \\
max. poletip field of quads., ${\rm B_{4\sigma}}$ [T]
                                        & 6 & 10 & 10 & 15 & 20 \\
max. full aperture of quad., ${\rm A_{\pm4\sigma}}$[cm]
                                        & 14 & 13 & 30 & 20 & 13 \\
${\rm \beta_{max} [km]}$               & 1.5 & 22 & 450 & 1100
                                        & 61 000 \\
final focus demagnification, $\sqrt{\beta_{\rm max}/\beta^*}$
                                        & 220 & 2200 & 19 000 & 31 000
                                       & 280 000 \\
\hline
\multicolumn{1}{|l|}{\bf synchrotron radiation parameters:} & & & & & \\
syn. E loss/turn [MeV]                 & 0.0008
                                        & 0.01 & 0.9 & 17 & 25 000 \\
syn. rad. power [kW]                   & 0.0002 & 0.13 & 0.4 & 400
                                        & 110 000 \\
syn. critical E [keV]                  & 0.0006 & 0.09 & 1.6 & 12 & 1700 \\
\hline
\multicolumn{1}{|l|}{\bf neutrino radiation parameters:} & & & & & \\
collider reference depth, D[m]           & 10 & 125 & 300 & 300 & 300 \\
ave. rad. dose in plane [mSv/yr]               & $3 \times 10^{-5}$
                                        & $9 \times 10^{-4}$
                                        & $9 \times 10^{-4}$
                                        & 0.66 & 6.7 \\
str. sect. length for 10x ave. rad.,
${\rm L_{x10}}$[m] & 1.9 & 1.3 & 1.1 & 1.0 & 2.4 \\
$\nu$ beam distance to surface [km]    & 11 & 40 & 62 & 62 & 62 \\
$\nu$ beam radius at surface [m]       & 24 & 8.4 & 3.3 & 1.3 & 0.13 \\ \hline

\end{tabular}
\label{specs}
\end{center}
\end{table}

\end{document}